\documentclass[showpacs,amsmath,amssymb,floatfix,prl,twocolumn]{revtex4}

\usepackage[pdftex]{graphicx}
\usepackage{amssymb,amsfonts,amsmath}
\usepackage{graphicx,amsmath}
\usepackage{float}

\begin{document}

\title{Quantized escape and formation of edge channels at high Landau levels.}
\author{A.D. Chepelianskii$^{(a,b)}$, J. Laidet$^{(c)}$, I. Farrer$^{(a)}$, D.A. Ritchie$^{(a)}$, K. Kono$^{(b)}$, H. Bouchiat$^{(c)}$\\
(a) Cavendish Laboratory, University of Cambridge, J J Thomson Avenue, Cambridge CB3 OHE, UK\\
(b) Low Temperature Physics Laboratory, RIKEN, Wako, Saitama 351-0198, Japan \\
(c) LPS, Univ. Paris-Sud, CNRS, UMR 8502, F-91405, Orsay, France\\
}


\begin{abstract}
We present nonlocal resistance measurements in an ultra high mobility two dimensional electron gas.
Our experiments show that even at weak magnetic fields classical guiding along edges leads
to a strong non local resistance on macroscopic distances. In this high Landau level regime the transport 
along edges is dissipative and can be controlled by the amplitude of the voltage drop along the edge.
We report resonances in the nonlocal transport 
as a function of this voltage that are interpreted as escape and formation of edge channels.
\end{abstract}

\maketitle

The investigation of nonlocal effects in electrical transport has provided
new insights on non classical conduction mechanisms. These effects
are responsible for the appearance of a potential difference across a region of the sample 
well outside of the classical current paths. They have been 
reported in conductors that exhibit quantum coherence \cite{Buttiker,BenoitWebb,Chandrasekhar}, ballistic transport \cite{vanHouten,Kim} 
or in the quantum Hall effect regime of a two dimensional electron gas \cite{Sacks,Geim,Shiraki}.
In the latter case the non local resistance appears due to the formation 
of edge channels that are isolated from the bulk and can carry the current 
to classically non accessible regions. The propagation of edge channels in this regime,
has attracted a significant interest due to their potential for quantum computation 
and interferometry \cite{Shtrikman,Altimiras,Shayegan,Bocquillon}. 
Here, using non local measurements 
we consider the opposite limit of high Landau levels where the bulk density of states 
is gap-less. We show that in this limit the exchange of charges between bulk and edge states
can be controlled by the voltage drop along the edges, which leads to the formation of resonances 
in the non linear transport that allows us to observe directly a quantization of edge channels
at high Landau levels.

\begin{figure}
\centerline{\includegraphics[clip=true,width=8cm]{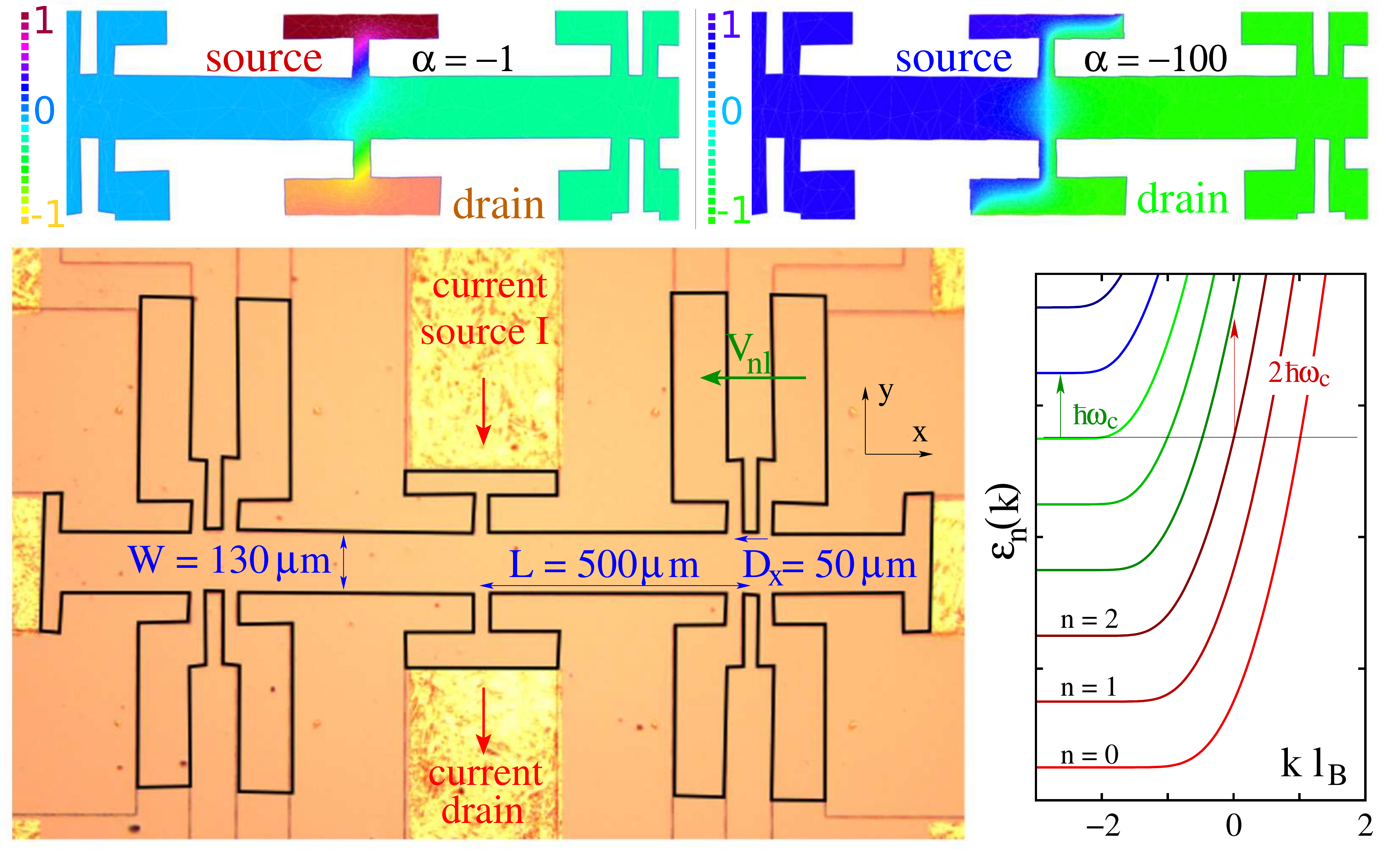}}
\caption{Sample geometry in our non local transport experiments. Arrows indicate the geometrical parameters in our 
experiment, the position of the source and drain electrodes and the electrodes across which the nonlocal potential 
drop $V_{nl}$ is measured. The nonlocal resistance is then defined as $R_{nl} = V_{nl}/I$.  
The closed black contour highlights the geometry of the domain used in our finite element simulations whose results are 
displayed in the top panels for $\alpha = \rho_{xy}/\rho_{xx} = -1$ and $\alpha = -100$,
the color/gray-scale level indicate the potential values (source/drain potentials are fixed to $\pm 1$);
the potential gradients are concentrated in the center of the sample. The curves on the 
right represent the dispersion relation $\epsilon_n(k)$ for edge states for a hard wall potential, $k$ is the wavenumber and 
$l_B$ is the magnetic length \cite{Avishai}.
}
\label{FigPhotoFem}
\end{figure}

\begin{figure}
\centerline{\includegraphics[clip=true,width=8cm]{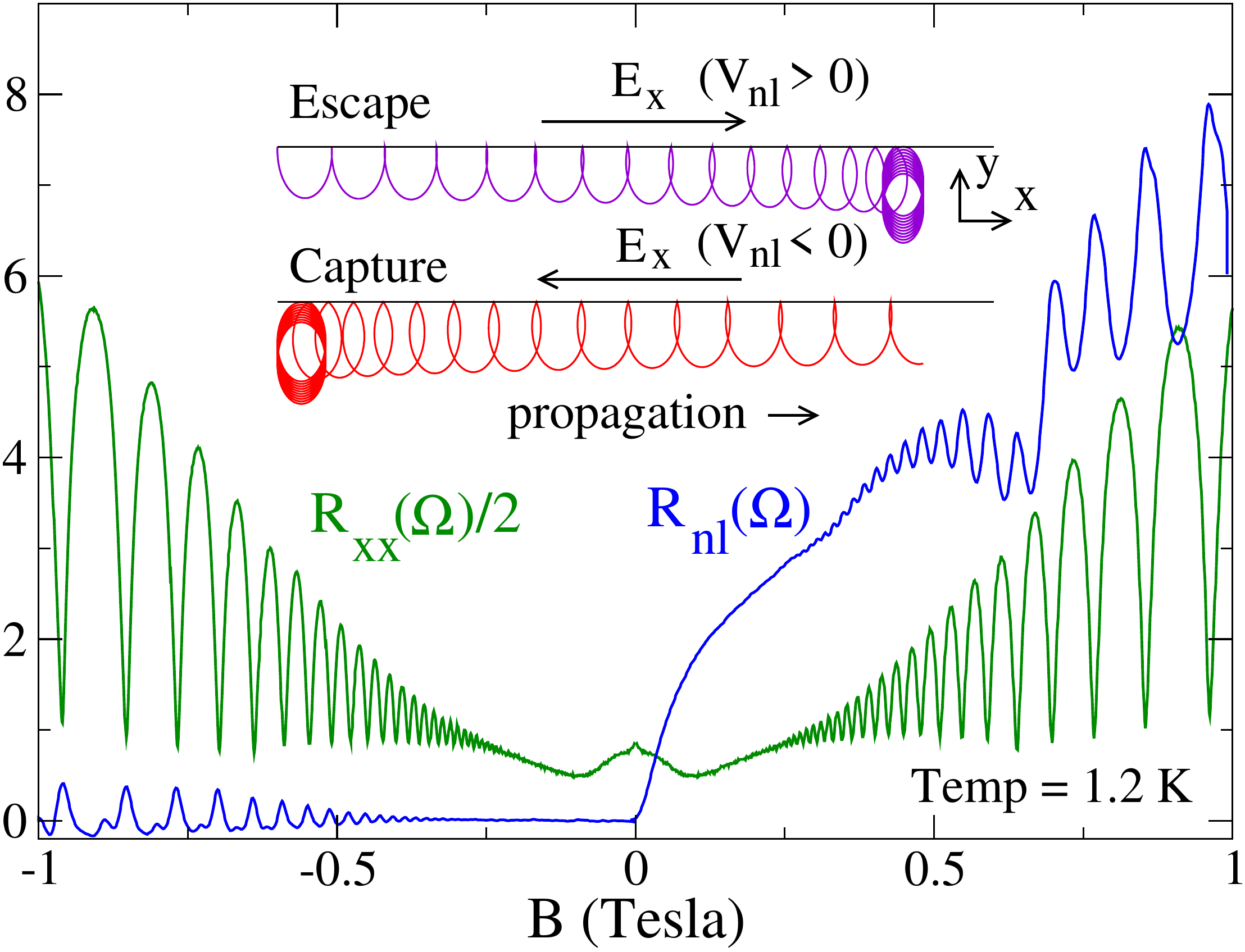}}
\caption{Dependence of the non local resistance $R_{nl}$ (as defined in Fig.~\ref{FigPhotoFem}) and of the longitudinal resistance $R_{xx} \simeq \rho_{xx}$ on the magnetic field $B$. The longitudinal resistance $R_{xx}$ is almost a symmetric function of $B$ whereas $R_{nl}(B)$ is strongly asymmetric and almost vanishes for $B < 0$. The insets illustrate typical classical electron orbits for a capture and an escape event due to the parallel electric field $E_x$ for $B > 0$ where electrons propagate along the upper edge in the positive $x$ direction. Capture occurs for $V_{nl} < 0$ and escape occurs for $V_{nl} > 0$.
}
\label{FigExpLin}
\end{figure}

We have investigated the magnetic field dependence of nonlocal transport in a $GaAs/Ga_{1-x}Al_x As$
2DEG with density $n_e \simeq 3.3\times10^{11}{\rm cm}^{-2}$, 
mobility $\mu \simeq 10^7\;{\rm cm^2/Vs}$ corresponding to transport time $\tau_{tr} \simeq 0.4\;{\rm ns}$ 
and a mean free path of $\ell_e = 100\;{\rm \mu m}$.
The Hall bar with a channel width $W = 100\;{\rm \mu m}$ was patterned using wet etching. The non local 
resistance $R_{nl}$ was measured in a geometry illustrated on Fig.~\ref{FigPhotoFem} where current was injected along the $y$ axis and the voltage was detected between two probes distant by $D_x \simeq 50\;{\rm \mu m}$ at a distance $L \gg W$ from 
the current injection points. 
The experimental data on Fig.~\ref{FigExpLin}, show that $R_{nl}$ 
exhibits an unusual dependence on magnetic field that in striking difference from $\rho_{xx}$ behavior. 
Indeed, in contrast to $\rho_{xx}(B)$, $R_{nl}(B)$ is a strongly asymmetric function of the magnetic field
that almost vanishes for negative magnetic fields and exhibits a sharp onset at low positive magnetic fields 
reaching a value of the order of $\rho_{xx}$ for $B > 0.1\;{\rm Tesla}$.

We have first checked whether this dependence can be explained using the continuum theory of a Hall bar. 
For this purpose, it is convenient to describe our sample as a 2DEG stripe, and to approximate the current 
injection leads by point-like sources. This stripe can be parametrized by complex numbers 
$z = x + i y$ with $y \in (0, W)$. 
The potential $V(z)$ created by a current source $I$ positioned at $x = x_0$ along the top/bottom edges
then reads $V_\pm(z) = R_\pm(z, x_0) I$ (plus/minus sign for top/bottom edge) where:
\begin{align}
R_\pm(z,x_0) = R_p( \exp(\frac{\pi z}{W}) \exp(\frac{-\pi x_0}{W}) \pm 1).
\label{eqRmoins}
\end{align}
The function $R_p$ gives the potential created by an unit current source located at the origin in the semi-infinite 2DEG half plane $y > 0$:
\begin{align}
R_p(z) &= \frac{\rho_{xx}}{\pi} \left( \log |z| + \alpha \arg z \right) 
\end{align}
where we introduced the notation $\alpha = \rho_{xy}/\rho_{xx}$. 

Subtracting these two expressions we find the potential $V = (R_+(z,0) - R_-(z,0)) I$ created by a current between 
point-like source and drain located opposite to each other along the channel.
Using these equations for the particular case of the potential generated along the top edge $z = i W$, 
far from the sources $|x| \gg W$, we find the following expression for the non local resistance: 
\begin{align}
R_{nl} = \frac{2 \rho_{xx} D_x}{W} \exp\left( -\frac{\pi L}{W} \right)
\label{eq:Rnlcontinuum}
\end{align}
where $D_x$ is the spacing between the voltage probes and $L$ is their distance from the source along the channel
(for simplicity we have assumed $D_x \ll W$).
The geometrical parameters in our experiment are $L \simeq 500\;{\rm \mu m}$, $W \simeq 130\;{\rm \mu m}$ 
and $D_x \simeq 50\;{\rm \mu m}$ (see Fig.~\ref{FigPhotoFem}), which lead to a numerical estimation $R_{nl} \simeq 4.4 \times 10^{-6} \rho_{xx}$.
Thus according to this point source model the nonlocal resistance is proportional to $\rho_{xx}$ with an exponentially 
small damping factor which is independent of the magnetic field. This conclusion however is in strong disagreement
with the experimentally observed dependence. 
In order to check the validity of this analytical estimation in our more complex experimental geometry, 
we have performed a finite element simulation of the potential 
which (see Fig.~\ref{FigPhotoFem}) confirms the exponential decay of the field amplitudes away 
from the current polarization contacts. 



Thus even at small magnetic fields ($\le 0.1\;{\rm Tesla}$),  our experiments indicate a
large non local resistance that cannot be described within the continuum theory.  
Due to the macroscopic dimensions of our sample (channel width $W \simeq 130\;{\rm \mu m}$) 
quantum coherence effects cannot explain the origin of the non local resistance in our measurements.
An explanation relying solely on the formation of Landau levels is also unlikely since we observe
$R_{nl} \sim \rho_{xx}$ even at weak magnetic fields $B \simeq 0.1 {\rm Tesla}$ where Shubnikov-de Haas
oscillations are absent. 
We thus propose guiding along sample edges as a possible explanation for the observed behavior
and attempt to include the physics of skipping orbits within the continuum model.
The formation of skipping orbits occurs due to the bending of the Landau levels at the edge
of the 2DEG \cite{Avishai} which is represented on Fig.~\ref{FigExpLin}. It can lead to noticeable 
effects even when individual Landau levels are not resolved \cite{beenakker}.

In presence of skipping orbits, electrons can propagate along the edges before being injected 
into the bulk of the 2DEG. This gives rise to edge currents $I_+, I_-$ along the top and bottom edges 
of the sample. Due to the influence of disorder electrons will progressively detach from the edges causing 
a progressive drop of the edge current in the direction of propagation of the electrons. The drop 
in the current carried by the edges $dI_+/dx$ and $dI_-/dx$ creates a distributed current source for the bulk of 
the 2DEG. The equations Eqs.~(\ref{eqRmoins}) derived within the continuum model, allow us to find the 
potential created by this distributed current source: 
\begin{align}
V = -\int R_+(z,x) \frac{d I_+(x)}{d x} d x - \int R_-(z,x) \frac{d I_-(x)}{d x} d x 
\label{Vnonloc}
\end{align}
We will assume that the edge currents are non zero only in the direction of propagation of the electrons 
and decay exponentially with a characteristic length-scale $\lambda_e$ that we will call the mean free path 
along edges, this leads to: 
\begin{align}
I_+(x) = I_-(-x) = s_B I e^{-s_B x / \lambda_e} \eta(s_B x) 
\label{Iedge}
\end{align}
where $s_B = \pm 1$ for positive/negative magnetic fields and $\eta$ is the Heaviside function. 
It is straightforward to check that the total current $-\int \frac{d I_+(x)}{d x} d x$, 
injected into the bulk 2DEG from the top electrode is $I$. 
Assuming $|\alpha| \gg 1$ and combining Eqs.~(\ref{Vnonloc}),(\ref{Iedge}) we find the following approximation for the non local resistance: 
\begin{align}
R_{nl} = \rho_{xy} \frac{W}{\lambda_e} \exp\left( -\frac{\pi L}{\lambda_e} \right) \eta(s_B) 
\label{eq:Rnledge}
\end{align}
Since this equation was derived assuming that electrons were guided only in one direction,
it predicts a vanishing non local resistance for negative magnetic fields,
in qualitative agreement with the experiment.
We note however that for $B < -0.5\;{\rm Tesla}$, 
a finite non local resistance of oscillating sign appears that is not expected within this model. 
A possible origin of this effect, could be from electrons that are recaptured by the edges 
after moving through the bulk of the samples and that are not accounted for in the present model.
At positive magnetic fields, this equation can be used to estimate $\lambda_e$ from the experimental data,
which yields for $B \ge 0.1\;{\rm Tesla}$, $\lambda_e \simeq 90\;{\rm \mu m}$.
Weak variations of $\lambda_e$ as a function of the magnetic field (at most $10\%$)  can explain
the presence of Shubnikov-de Haas oscillations in $R_{nl}(B)$. 
We note that the obtained value $\lambda_e$ is very close to mean free path in the sample $\ell_e \simeq 100\;{\rm \mu m}$.

\begin{figure}
\centerline{\includegraphics[clip=true,width=8.5cm]{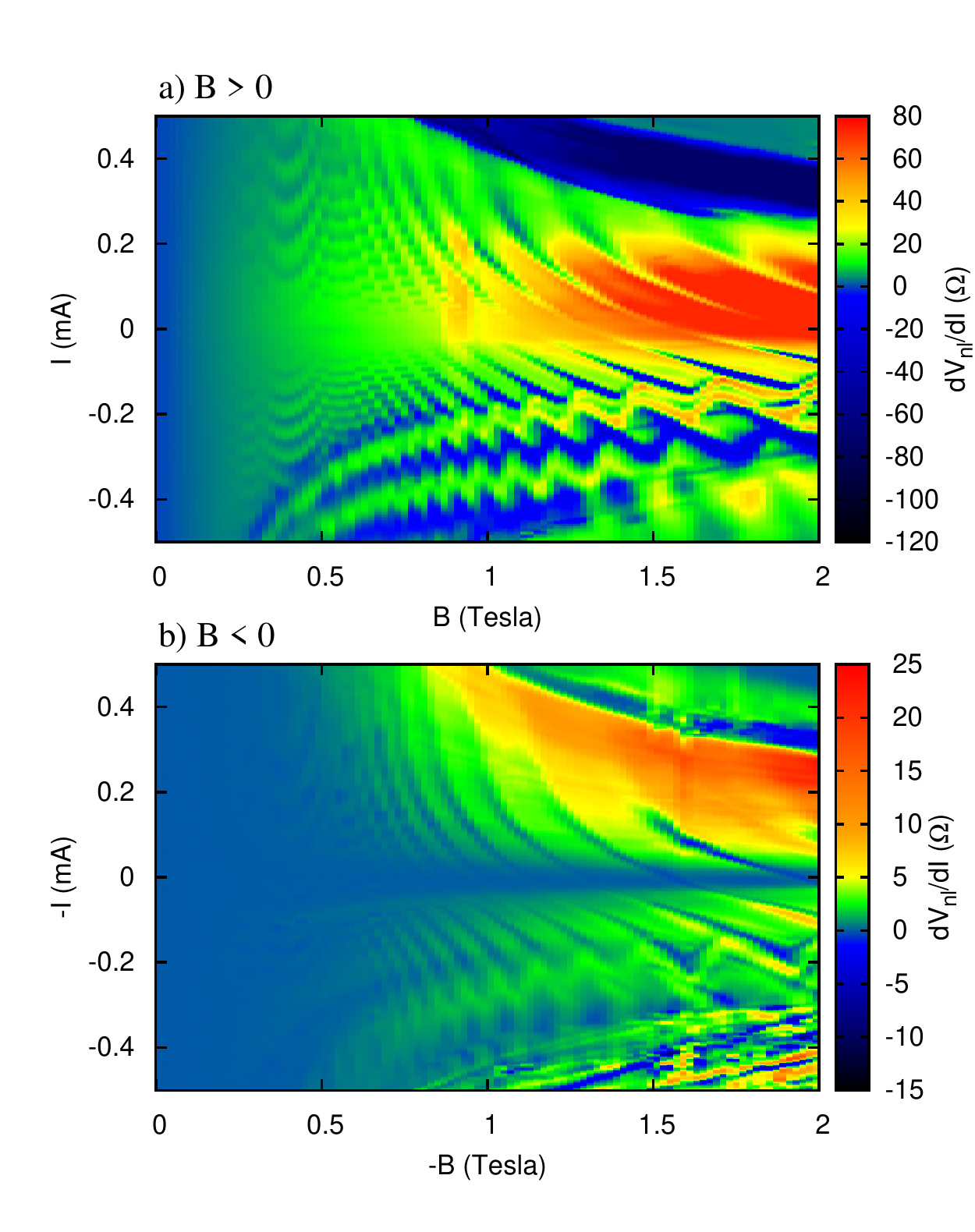}}
\caption{Dependence of the differential non local resistance $d V_{nl}/d I$ 
on magnetic field $B$ and on the DC current amplitude $I$ for positive and negative magnetic fields (top/bottom panels). 
The data at negative magnetic fields is displayed as a function of $-B$ and $-I$. Temperature was $1.2\;{\rm K}$.
}
\label{FigdVdIMega}
\end{figure}

Even if the proposed model describes qualitatively the observed non local resistance,
it is based on a phenomenological assumption on the distribution of the edge currents $I_e(x)$,
and a microscopic theory is needed to determine self consistently,
the potential inside the device and the distribution of the edge currents.
Several approaches have been proposed to treat the interaction between bulk and 
edge transport in the quantum limit at low filling factors 
\cite{Sacks,Svoboda,Brinkman,Johnson} and do not directly apply to the present case.
Indeed, the propagation along edge channels has mainly been studied 
at an integer Quantum Hall effect plateau, where the transport is non dissipative $R_{xx} = 0$
and a gap in the density of states opens in the bulk \cite{Altimiras,Bocquillon}.

\begin{figure}
\centerline{\includegraphics[clip=true,width=7cm]{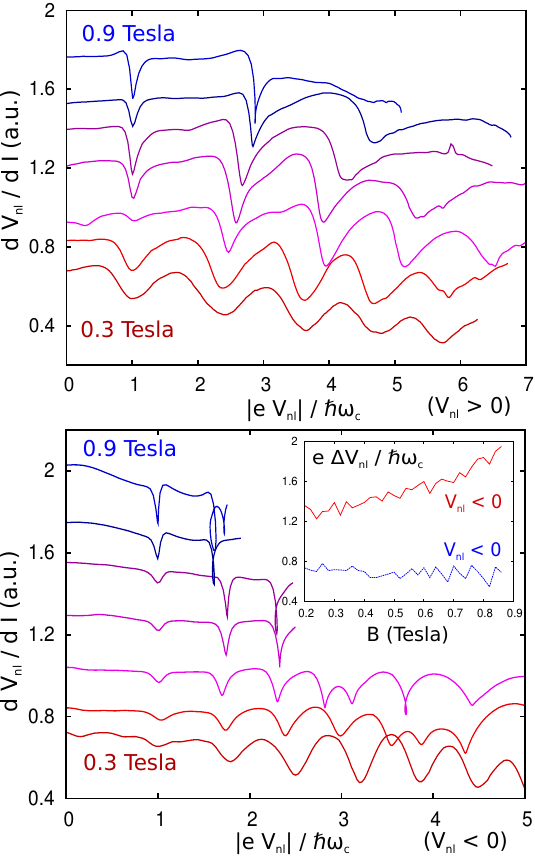}}
\caption{Dependence of the differential non local resistance $d V_{nl}/d I$ (in arbitrary units) 
on the dimensionless quantity $x = |e V_{nl}|/\hbar \omega_c$ at magnetic fields between $0.3$ and $0.9\;{\rm Tesla}$, 
a voltage offset was applied to fix the position of the first resolved peak at $x = 1$. 
The period, of the oscillations is plotted as a function of magnetic field in the inset for positive and negative $V_{nl}$. 
It corresponds to the distance between the first resolved peaks at magnetic fields where only a few oscillations could be resolved. 
}
\label{FigdVdIvsV}
\end{figure}

In our case, due to the low magnetic fields the gap is not present and electrons can escape to the bulk
or on the contrary approach towards the edge.
To look for signatures of the escape and creation of edge channels,
we have measured the differential nonlocal resistance $d V_{nl}/d I$ as a function 
of magnetic field and DC excitation current $I$.
At positive magnetic fields, when the potential $V_{nl}$ is positive electrons lose an energy $|e| V_{nl}$ as 
they cross the separation distance between the voltage probes, thus some electrons will escape from the edge,
because their Larmor radius becomes smaller as they propagate. 
If the potential $V_{nl}$ is negative, electrons in the bulk will 
tend to drift towards the edge under the action of the electric field $E_x = V_{nl}/D_x$ and new edge channels may be formed.
The typical trajectories for a capture and an escape event are represented on Fig.~\ref{FigExpLin}.
We thus expect that the transport properties along the edge will strongly depend on the sign of $V_{nl}$.

In agreement with our heuristic arguments, the experimental results displayed on Fig.~\ref{FigdVdIMega} 
exhibit a striking asymmetry between positive and negative currents. For positive currents 
(at $B \ge 0.5\;{\rm Tesla}$) we measure positive $d V_{nl}/d I$ for $I > 0$ whereas 
for $I < 0$, $d V_{nl}/d I$ drops and exhibits sharp oscillations around zero.
To ensure that this difference is not related 
to some asymmetry of the sample, we have also measured $d V_{nl}/d I$ at negative magnetic fields. 
Except for the region around $I = 0$ where the differential resistance almost vanishes in agreement with our guiding 
model, we find that after the transformation $I \rightarrow -I$, 
results are very similar to those obtained at $B > 0$.
This observation confirms that our findings cannot be attributed to a geometrical asymmetry 
which would not depend on the sign of the magnetic field. 
To understand, the origin the approximate symmetry observed in Fig.~\ref{FigdVdIMega}, we note that a mirror symmetry around the Hall bar channel changes, 
$I \rightarrow -I$ and $B \rightarrow -B$ and inter-exchanges top and bottom edges.
The non local voltage across the bottom edge is therefore expected to be $V_{nl}(-B, -I)$, 
the electrons emitted from the bottom edge can then be recaptured 
at the top edge where $d V_{nl}/d I$ is measured, giving a contribution proportional to $d V_{nl}/d I$ at $B > 0$ and current $-I$
damped by the propagation through the bulk. Hence from now on we will focus on the analysis of the data 
obtained at $B > 0$.

The dependence on $I$ displayed in Fig.~\ref{FigdVdIMega}, exhibits several intriguing features.
To gain an understanding on their physical origin, we will concentrate on the region of weak magnetic fields
($B$ between $0.2$ and $0.9\;{\rm Tesla}$). In this region $d V_{nl}/d I$ exhibits smooth oscillations as a function 
of $I$, integrating on current we find the dependence $V_{nl}(I)$ and display the differential resistance 
as a function of $e V_{nl}/\hbar \omega_c$ (see Fig.~\ref{FigdVdIvsV}), where $\hbar \omega_c$ is the spacing between 
Landau levels. After this transformation, the origin of the oscillations becomes more conspicuous,
at low magnetic fields an oscillation $d V_{nl}/d I$ occurs whenever $e V_{nl}$ is changed by approximately $\hbar \omega_c$.
The dependence of the period $\Delta V_{nl}$ on the magnetic field is displayed on the inset of Fig.~\ref{FigdVdIvsV}.
For $V_{nl} < 0$ where we expect formation of new edge channels due to drift of bulk electrons towards the edge,
$\Delta V_{nl}$ is almost equal to $\hbar \omega_c/e$ (we attribute the $20\%$ difference, 
to the aspect ratio between the distance between voltage probes and their width). 
However for $V_{nl} > 0$, when electrons loose energy as they propagate and edge channels progressively 
escape to the bulk, the ratio $e \Delta V_{nl}/\hbar \omega_c$ progressively increases with magnetic field.
Our interpretation is that at $V_{nl} < 0$, we are probing the outermost edge channels that have an
energy spacing close to $\hbar \omega_c$, while for $V_{nl} > 0$ edge channels escape progressively 
and only the inner channels with an energy spacing larger than $\hbar \omega_c$ are still propagating
(see level diagram in Fig.~\ref{FigPhotoFem}).

As the magnetic field increases the following trends can be noted, for $I > 0$ the smooth oscillations 
develop into sharp resonances at certain values of $V_{nl}$, while for negative currents $d V_{nl}/d I$ start to change
sign as function of $I$ rendering our analysis as a function of $V_{nl}$ impossible. 
Experiments with a larger separation between voltage 
probes $D_x \simeq 500\;{\rm \mu m}$, did not display the described oscillation and resonances,
which suggests that their observation is possible only when $D_x$ is smaller than the mean free path.
In a control sample with wide voltage probes of around $300\;{\rm \mu} m$,  
a zero differential resistance plateau was observed at $I < 0$,
indicating that in this regime the electrostatic potential oscillates as a function of 
the distance along the edge and averages to zero when voltage is measured on a large length scale \cite{supp}. 
A vanishing differential resistance has previously been reported in local measurement geometries \cite{bykov2007,zudov2011}
where bulk and edge contributions are intermixed; our experiments show that a zero differential resistance state 
can be created by edge effects alone. 
Additional experimental and theoretical investigations are needed to fully understand 
edge transport at high Landau levels in the nonlinear regime. It would also be interesting to perform similar experiments 
under microwave irradiation where stabilization of edge channels is expected \cite{toulouse}
and where non local effects can also be present \cite{Kono2}.

To summarize, we have demonstrated through non local resistance measurements that guiding effects can strongly 
modify the potential distribution in ultra high mobility samples even in the limit of weak magnetic fields 
$B \le 0.1\;{\rm Tesla}$. In the linear transport regime, our observations are consistent 
with a spreading of the distribution of the current source in the direction of propagation along edges.
As opposed to the quantum Hall regime where transport in the bulk is suppressed, an exchange between
edge and bulk conduction paths takes place in our experiments. We show that this exchange can be controlled 
by the amplitude of the potential drop along the edge. Additional edge channels can be formed if the electrons gain energy as they 
propagate along the edge, in the opposite case when electron loose energy the edge channels can escape to the bulk. 
We propose that oscillations in non linear transport when the amplitude of the voltage drop along 
the edge is changed by the spacing between Landau levels are a signature of quantized escape and formation 
of edge channels. Thus edge transport in the limit of high filling factors allows to explore a rich 
physical regime that may have deep implications in our understanding of electron transport 
in ultra clean systems. We thank M. Polianski and I.A. Dmitriev for fruitful discussions and acknowledge support from St. Catharine college and Toshiba Research Europe.

\clearpage 


\end{document}


\title{Quantized escape and formation of edge channels at high Landau levels : supplementary materials}
\author{A.D. Chepelianskii$^{(a,b)}$, J. Laidet$^{(c)}$, I. Farrer$^{(a)}$, D.A. Ritchie$^{(a)}$, K. Kono$^{(b)}$, H. Bouchiat$^{(c)}$\\
(a) Cavendish Laboratory, University of Cambridge, J J Thomson Avenue, Cambridge CB3 OHE, UK\\
(b) Low Temperature Physics Laboratory, RIKEN, Wako, Saitama 351-0198, Japan \\
(c) LPS, Univ. Paris-Sud, CNRS, UMR 8502, F-91405, Orsay, France\\
}


\begin{abstract}
The supplementary materials describe the evolution of the non-linear transport in non-local geometry,
from a series of resonances corresponding to escape and creation of edge channels (presented in the main article)
towards a zero differential resistance state when the voltage drop is measured on length scales much 
larger than the mean free path. The magnetic field ($B \rightarrow -B$) and DC current ($I \rightarrow -I$) symmetry
properties of the reported zero-differential state strongly supports its edge transport origin.
Finally we provide a more detailed derivation for the equations of the continuum theory.
\end{abstract}

\maketitle

\section{I. Non local differential resistance with distant voltage probes}

\begin{figure}
\centerline{
\includegraphics[clip=true,width=8.5cm]{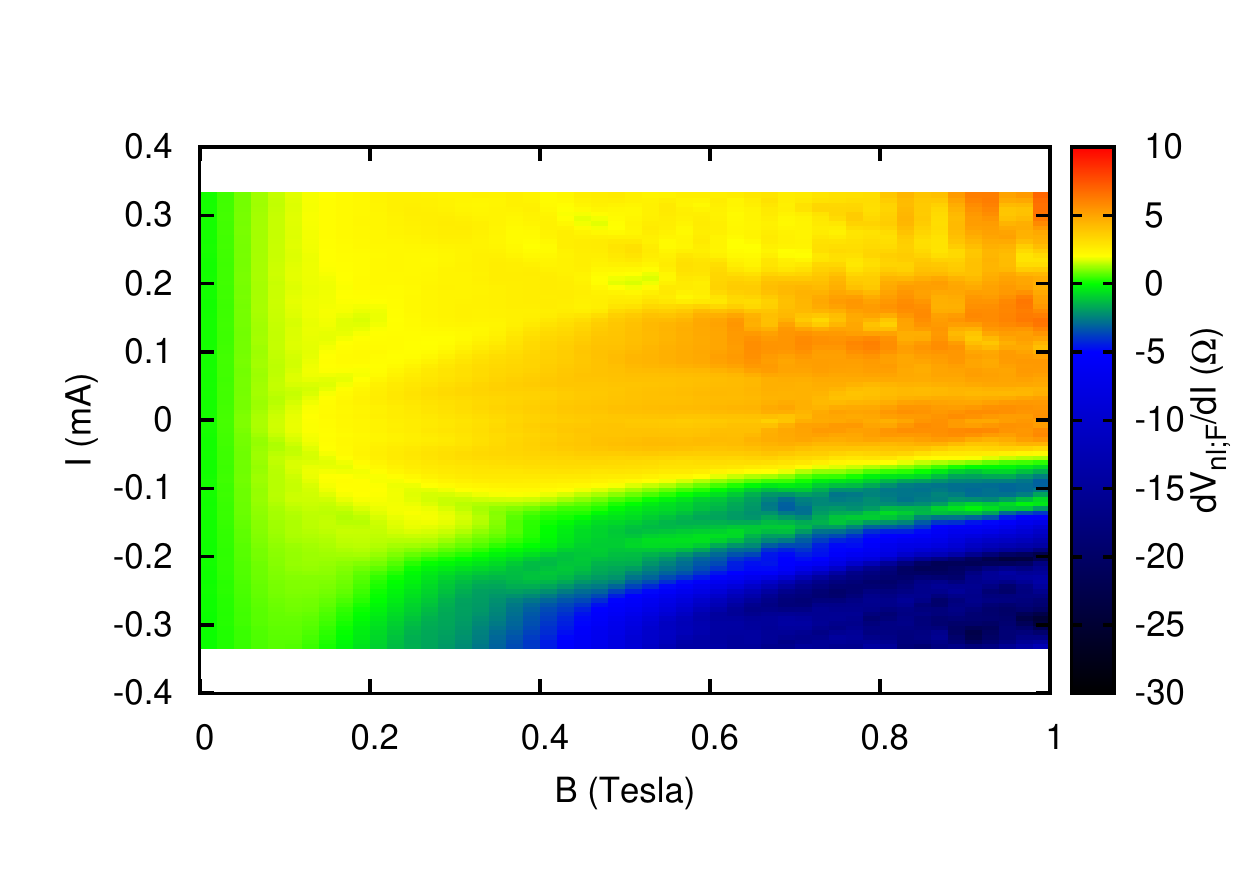}}
\caption{Dependence of the non local differential resistance $d V_{nl;F}/d I$ on magnetic field and DC current amplitude, this quantity was measured in a geometry 
where the separation between voltage probes was $D_x \simeq 500\;{\rm \mu m}$ on the $\mu = 10^7\;{\rm cm^2/Vs}$ sample from the main text. Temperature was $T = 1.2\;{\rm K}$.
}
\label{Supp1}
\end{figure}

We have measured the non local differential resistance (NLDR) $d V_{nl;F}/d I$ 
in a geometry where the voltage probes were separated by a distance $D_x = 500\;{\rm \mu m}$ 
larger than the mean free path $\ell_e = 100\;{\rm \mu m}$ in the sample.
The experiment was performed on the same sample as in the main text but with a different arrangement of 
voltage and current probes, the current sources were located $500\;{\rm \mu m}$ away from the voltage probes.
In the linear response regime the dependence of $R_{nl;F} = d V_{nl;F}/d I(I = 0)$ on the magnetic field, 
was very similar to the data shown on Fig.~2 (from main article).
The quantity $R_{nl:F}$ was finite for positive magnetic fields and almost vanished for $B < 0$.
The dependence of $d V_{nl;F}/d I$ on the magnetic field 
$B$ and on the DC current amplitude $I$ is represented on Fig.~\ref{Supp1} for $B > 0$.
The oscillating features as function of the DC current $I$ are not resolved
contrarily to measurements where $D_x$ was smaller than the mean free path (see data on Fig.~3 from main article).
Negative values of $d V_{nl;F}/d I$ at negative current $I$ are still observed in this geometry.

\section{II. Formation of zero differential resistance states in a macroscopic sample}

\begin{figure}
\centerline{
\includegraphics[clip=true,width=8.5cm]{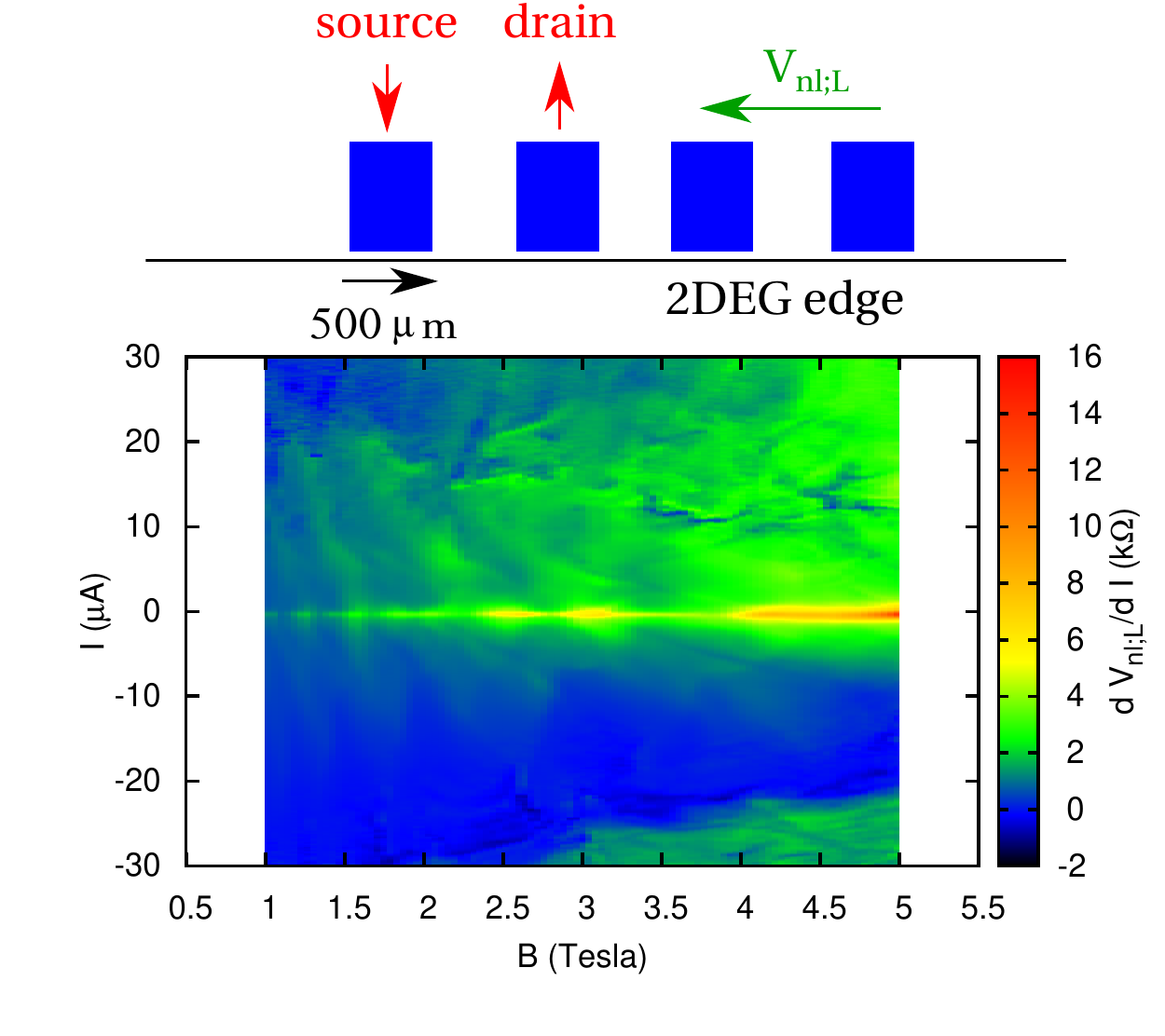}}
\caption{Dependence of NLDR $d V_{nl;L}/d I$ on magnetic field and DC current amplitude for the $\mu = 3 \times 10^6\;{\rm cm^2/Vs}$ mobility sample. NLDR was measured in the geometry sketched in the top panel, temperature was $T = 0.3\;{\rm K}$.}
\label{Supp2}
\end{figure}

We have also studied NLDR in a macroscopic geometry with geometrical parameters larger than the mean free path.
The sample was made in a lower mobility 2DEG, with mobility $\mu = 3 \times 10^6\;{\rm cm^2/Vs}$ 
and a carrier density of $n_e = 3.2\;\times 10^{11} {\rm cm}^{-2}$. 
The geometry of the measurement is sketched in in Fig.~\ref{Supp2}.
This figures summarizes our results on the NLDR in this sample,
for positive magnetic fields for which the non local resistance is non-vanishing. 

The strong asymmetry between positive and negative currents
is also observed in this lower mobility 2DEG, however the characteristic magnetic field where 
the asymmetry appears is around a factor three stronger as compared to the $\mu = 10^7\;{\rm cm^2/Vs}$ 
sample, this difference is consistent with the ratio between the mobilities of the two samples. 
As in Fig.~\ref{Supp1}, the separation between the voltage probes 
was larger than the mean free path $\ell_e = 30\;{\rm \mu m}$
and the oscillations as a function of the DC current cannot be resolved. 
However, in the present experiment NLDR is almost zero in a large region of negative currents 
which contrasts with previous data where NLDR could be negative for $I < 0$
(see Fig.~2 from the main article and Fig.~\ref{Supp1}).

\begin{figure}
\centerline{
\includegraphics[clip=true,width=9cm]{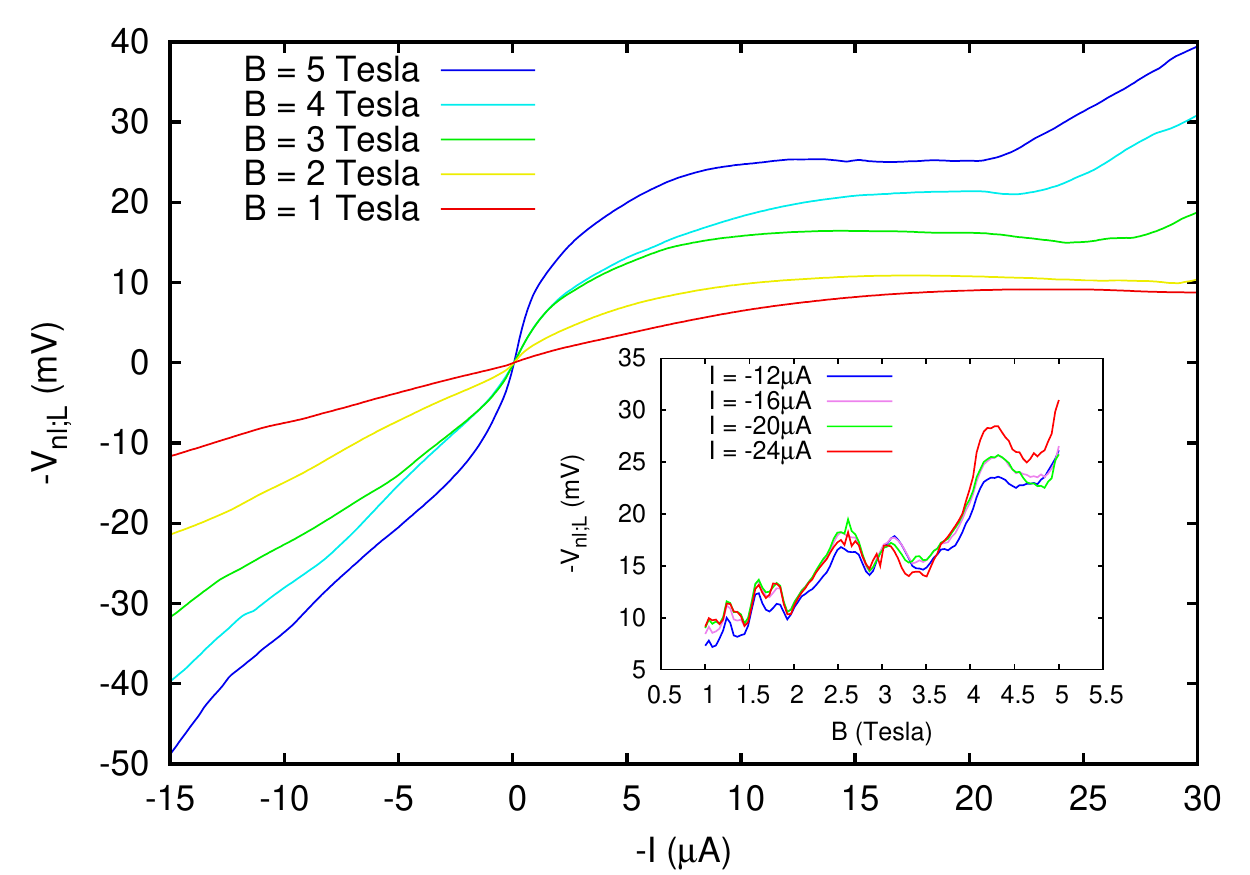}}
\caption{Non local voltage/current characteristics $V_{nl;L}(I)$ for the $\mu = 3 \times 10^6\;{\rm cm^2/Vs}$ mobility sample at 
several magnetic fields (for resemblance with data from ZDRS experiments in local geometries, 
we have shown $-V_{nl;L}$ as function of $-I$ in this figure). The inset shows the voltage as a function of magnetic field for several 
currents inside the plateau regime. Temperature was $T = 0.3\;{\rm K}$.}
\label{Supp3}
\end{figure}

In order to highlight the presence of a zero differential resistance state (ZDRS),
we have calculated the dependence of $V_{nl;L}$ on current by integrating the experimental differential resistance data.
The results obtained after this procedure are represented on Fig.~\ref{Supp3} which shows that 
the voltage $V_{nl;L}$ exhibits a plateau at negative $I$ where it is almost independent on current in a wide 
range of magnetic fields while for positive currents the voltage dependence is almost ohmic. 
The inset in Fig.~\ref{Supp3}, shows the dependence of the voltage on the magnetic field for several 
values of current inside the ZDRS plateau. These results confirm that the voltage saturates to a constant value 
independent on current in this regime, the value of the saturation voltage grows almost linearly with magnetic field
with weak oscillations that are probably related to the Shubnikov-de Haas oscillations in the longitudinal resistance.

The observed zero-differential state possesses the symmetry of an edge effect. It appears only for 
the sign of magnetic field which ensures guiding towards the voltage probe electrodes 
from the distant current sources, and for a specific sign of the DC current that creates a voltage drop 
along the edge tending to stabilize propagation along edges. Therefore it seems likely that an 
edge transport related mechanism is leading to the formation of ZDRS in this case. 
On the higher mobility sample where the dimension of the voltage probes were smaller than the mean free path,
negative values of NLDR were observed (see Fig.~3 from main article and Fig.~\ref{Supp1}), this suggests 
that ZDRS is formed due to the clamping of the potential on large length scales by the voltage probe electrodes. 
On the contrary, if the electrodes are not invasive the potential exhibits sharp variations whenever 
the energy of the electrons propagating along the edge is changed by an amount close to $\hbar \omega_c$ (see main text).
These voltage oscillations are probably indicative of a spatially modulated charge density distribution,
and could explain the observation of oscillating/negative differential resistances in our experiments. 
It would be highly interesting to understand the role played by the edge mediated ZDRS mechanism
in ZDRS experiments realized in the conventional longitudinal resistance measurement geometry.
However due to the absence of a reliable theoretical framework to describe the edge effects reported in this article, 
it is not possible to estimate the amplitude of their contribution in the measurement of longitudinal resistance.

\section{III. Continuum theory}

In this section we provide a mode detailed derivation of formulas from continuum theory that we used  
in the main article.

We start our calculations from the potential created by a point source of current I located at z = 0 in a semi-infinite two dimensional electron gas. It is convenient to represent points in the 2DEG as complex numbers z = x + iy where (x, y) are the point Cartesian coordinates, and the half plane fills the space $y > 0$. In this case we find the potential $V_p(z) = R_p(z) I$ with:
\begin{align}
R_p(z) &= \frac{\rho_{xx}}{\pi} \left( \log |z| + \alpha \arg z \right) 
\end{align}
where we have introduced the Hall angle $\alpha = \frac{\rho_{xx}}{\rho_{xy}}$.

A stripe geometry described by $z = x + i y$ with $y \in (0, W)$ 
can be mapped onto this half plane using the conformal mapping 
$z′ = \exp\left(\frac{\pi z}{W}\right)$. This allows to
find the potential $V_-(z, x_0) = R_-(z,x_0) I$ created by a point source located on
the bottom edge of the stripe at $z = x_0$ ($x_0$ real):
\begin{align}
V_-(z, x_0) = R_p( \exp(\frac{\pi z}{W}) \exp(\frac{-\pi x_0}{W}) - 1) I.
\end{align}

The potential $V_+(z, x_0) = R_+(z,x_0) I$ created by a source on the top edge of the
stripe at $z = x_0 + i W$ reads:
\begin{align}
V_+(z, x_0) = R_p( \exp(\frac{\pi z}{W}) \exp(\frac{-\pi x_0}{W}) + 1) I.
\end{align}

Subtracting these two expressions we find the potential
$V = V_+(z, 0) − V_−(z, 0)$ created by a current between
point-like sources and drains located opposite to each
other along the channel (respectively at $z = i W$ and
$z = 0$). For the particular case of the potential generated
along the top edge $y = i W$ , far from the sources $|x| \gg W$,
we find the following expression:
\begin{align}
V(x) = \frac{2}{\pi} I \rho_{xx} \exp\left( \frac{-\pi |x|}{W} \right) - \rho_{xy} I \eta(-x)
\end{align}

where $\eta(x)$ is the Heaviside function. This gives the expression for the nonlocal resistance given 
in the main text:
\begin{align}
R_{nl} = \frac{2 \rho_{xx} D_x}{W} \exp\left( -\frac{\pi L}{W} \right)
\end{align}

where $D_x$ is the spacing between the voltage probes and
$L$ is their distance from the source along the channel
(for simplicity we have assumed $D_x \ll W$).
